\documentclass[conference]{IEEEtran}
\IEEEoverridecommandlockouts
\usepackage{balance}
\usepackage{cite}
\usepackage{amsmath,amssymb,amsfonts}
\usepackage{algorithmic}
\usepackage{graphicx}
\usepackage{textcomp}
\usepackage[multiple]{footmisc}
\usepackage{multirow}
\usepackage{listings}
\usepackage{subfig}
\usepackage{makecell}
\usepackage[table,xcdraw]{xcolor}

\usepackage[pdftex]{hyperref}
\usepackage{tikz}

\newcommand\copyrighttext{%
    \footnotesize \textcopyright 2022 IEEE. Personal use of this material is permitted.
    Permission from IEEE must be obtained for all other uses, in any current or future
    media, including reprinting/republishing this material for advertising or promotional
    purposes, creating new collective works, for resale or redistribution to servers or
    lists, or reuse of any copyrighted component of this work in other works.
    DOI: \href{https://doi.org/10.1109/ICDMW58026.2022.00142}{https://doi.org/10.1109/ICDMW58026.2022.00142}}
\newcommand\copyrightnotice{%
    \begin{tikzpicture}[remember picture,overlay]
        \node[anchor=south,yshift=10pt] at (current page.south) {\fbox{\parbox{\dimexpr\textwidth-\fboxsep-\fboxrule\relax}{\copyrighttext}}};
    \end{tikzpicture}%
}

\def\BibTeX{{\rm B\kern-.05em{\sc i\kern-.025em b}\kern-.08em
    T\kern-.1667em\lower.7ex\hbox{E}\kern-.125emX}}

\begin{document}

\title{Macaw: The Machine Learning \\ Magnetometer Calibration Workflow}

\author{
\IEEEauthorblockN{
Jonathan Bader\IEEEauthorrefmark{1}\IEEEauthorrefmark{3},
Kevin Styp-Rekowski\IEEEauthorrefmark{1}\IEEEauthorrefmark{3},
Leon Doehler\IEEEauthorrefmark{3},
Soeren Becker\IEEEauthorrefmark{3},
Odej Kao\IEEEauthorrefmark{3},
}
\IEEEauthorblockA{
\IEEEauthorrefmark{3}
Technische Universit{\"a}t Berlin, Germany, \{firstname.lastname\}@tu-berlin.de}
}
\IEEEoverridecommandlockouts
\IEEEpubid{\makebox[\columnwidth]{*equal contribution \hfill} \hspace{\columnsep}\makebox[\columnwidth]{ }}
\maketitle
\copyrightnotice

\begin{abstract}

In Earth Systems Science, many complex data pipelines combine different data sources and apply data filtering and analysis steps.
Typically, such data analysis processes are historically grown and implemented with many sequentially executed scripts.
Scientific workflow management systems (SWMS) allow scientists to use their existing scripts and provide support for parallelization, reusability, monitoring, or failure handling.
However, many scientists still rely on their sequentially called scripts and do not profit from the out-of-the-box advantages a SWMS can provide.

In this work, we transform the data analysis processes of a Machine Learning-based approach to calibrate the platform magnetometers of non-dedicated satellites utilizing neural networks into a workflow called Macaw (MAgnetometer CAlibration Workflow).
We provide details on the workflow and the steps needed to port these scripts to a scientific workflow.
Our experimental evaluation compares the original sequential script executions on the original HPC cluster with our workflow implementation on a commodity cluster.
Our results show that through porting, our implementation decreased the allocated CPU hours by 50.2\% and the memory hours by 59.5\%, leading to significantly less resource wastage.
Further, through parallelizing single tasks, we reduced the runtime by 17.5\%.

\end{abstract}


    \section{Introduction}\label{sec:INTRO}
    Scientists frequently have to analyze enormous amounts of data that can easily exceed terabytes of input~\cite{heidsieck2019adaptive, vivian2017toil}.
Examples of such data analysis challenges are abundant in many scientific domains such as bioinformatics, where many sequences need to be examined in parallel~\cite{garcia2020sarek,yates2021reproducible, bader2021tarema, baderLotaruLocallyEstimating2022, baderReshi2022IPCCC}, in Earth observation~\cite{rettelbach2021quantitative,rettelbachGraphsSSDBM, lehmann2021force}, where hundreds of high-resolution pictures are analyzed, or in material science, where many molecules are analyzed in parallel~\cite{schaarschmidt2021workflow, stein2019progress}. 

The data processing is frequently executed on one of three types of infrastructures, a workstation (powerful personal computer), a commodity cluster (ensemble of multiple independent computers with commodity hardware, interconnected via a commodity communication network), or a High-Performance Computing (HPC) cluster (multiple independent high-performance computers interconnected via a very fast interconnect, e.g., Infiniband or Omni-Path)~\cite{regassa2022harvesting, kennedy2019parallel}.
Different infrastructure types require different usages. 
While a scientist can simply run the data processing scripts on his workstation without any changes, the usage of a commodity cluster or a HPC cluster might require code adaptions, e.g.,  the connection to a resource manager.

The problem of different infrastructure types is further aggravated since scientists frequently move their data between different compute infrastructures.
To provide a reproducible data pipeline that can be executed on various infrastructures with different resource managers, scientific workflow management systems (SWMS) can be used.
SWMS like Nextflow~\cite{nextflow}, Pegasus~\cite{deelman2019evolution}, or Snakemake~\cite{molder2021sustainable} claim to enable reproducible data analysis pipelines, providing out of the box parallelization, monitoring, and failure handling~\cite{lehmann2021force}.
They are provisioning adapters for different resource managers like Slurm~\cite{yoo2003slurm} or Kubernetes\footnote{\href{https://kubernetes.io/}{https://kubernetes.io/}} to support the execution on different clusters.
Further, lightweight virtualization technologies like Docker are used to enable reproducibility.
In addition, with SWMS, scientists can easily reuse their existing scripts and link them to a data analysis workflow instead of significantly rewriting their existing code.

In our paper, we present Macaw, a machine learning magnetometer calibration workflow.
The geomagnetic data for the calibration job is gathered from satellite missions and can be used to model and describe different components of the geomagnetic field, e.g., the core, the large-scale magnetospheric, and the lithospheric field~\cite{finlay2020chaos,baerenzung2020kalmag}.
The magnetic measurements of platform magnetometers from non-dedicated satellites are calibrated post-launch to a lower residual level which yields scientific value in modeling geoscientific phenomena by publishing the calibrated datasets. 
This calibration is done with a Machine Learning (ML) model trained for the regression task of reducing artificially introduced disturbances from the satellite system to the platform magnetometer readings.

We ported Macaw from the originally sequentially called scripts to the well-known workflow engine Nextflow~\cite{nextflow} in order to enable parallelization and a reproducible execution that can cope with errors.
This paper reports on the process of transforming the original implementation to the scientific workflow and the associated possible improvements during porting.
For our evaluation, we ran the original scripts on the initial HPC cluster and compared this execution with our workflow implementation on a commodity cluster, by leveraging evaluation metrics such as workflow runtime and resource utilization.

Overall, our workflow implementation was able to decrease the allocated CPU hours by 50.2\%, the allocated memory hours by 59.5\%, and the makespan (time between the start of the first task and the end of the last task) by 17.5\%.

\emph{Outline}. The remainder of the paper is structured as follows.
Section~\ref{sec:WORKFLOW} describes our original data analysis process.
Section~\ref{sec:WORKFLOW_IMPL} explains our workflow implementation in detail and highlights the challenges and adaptions necessary.
Section~\ref{sec:EVAL} evaluates our workflow and compares it to the original implementation.
Section~\ref{sec:DISCUSSION} discusses the results.
Section~\ref{sec:CONCLUSION} summarizes and concludes our paper.

    \section{Satellite Calibration Pipeline}\label{sec:WORKFLOW}
    
For this work, the magnetometer data of the ESA-operated mission GOCE (Gravity field and steady-state ocean circulation explorer) has been analyzed and calibrated.
This is a non-dedicated satellite, as the primary goal of the mission is not the acquisition of geomagnetic measurements, but it still carries a platform magnetometer for navigation and attitude control like many other satellites.
The original goals for these missions are manifold, e.g., image acquisition, geodetic, or climate studies \cite{borgeaud2015status}.
These platform magnetometers are part of the satellite body and are only roughly calibrated which is sufficient for their task.
The goal of the post-launch calibration, that is also described in more detail in \cite{styp2022machine} and \cite{michaelis2022geomagnetic}, is to correct for artificial disturbances to the measurements introduced by other satellite systems by comparison to a reference model.
This calibration of platform magnetometers yields the possibility of enlarging existing geomagnetic datasets.

Correcting for these artificial disturbances is a regression task as a multitude of information is collected on the satellite and included in the correction of the introduced artificial signals corresponding to different measured features of the satellite or its systems.
The measured data of the satellite needs to be prepared in different ways. 
As the data comes from different sources like raw magnetometer measurements, housekeeping data as well as telemetry data collected on the satellite, but also external geomagnetic field models or geomagnetic indices, the data needs to be interpolated and aligned to the same timestamps.
With this data, the regression model is able to correct for artificial disturbances introduced into the measurements by currents, temperatures, system activations, or other properties of the satellite system.
Then, the data can be merged and preprocessed with a variety of goals.
Data gaps occurring at times are corrected by a filling strategy utilizing the mean of available measurements.
If such gaps occur in critical measurements like the magnetometer or positional information, the data is not considered.
First, only geomagnetic quiet data is selected to model the satellite system itself instead of external natural phenomena, but also outliers are filtered away.
Afterward, the data is scaled and shuffled for the machine learning pipeline.

\begin{figure}[t]
\centering
    \includegraphics[width=1\columnwidth]{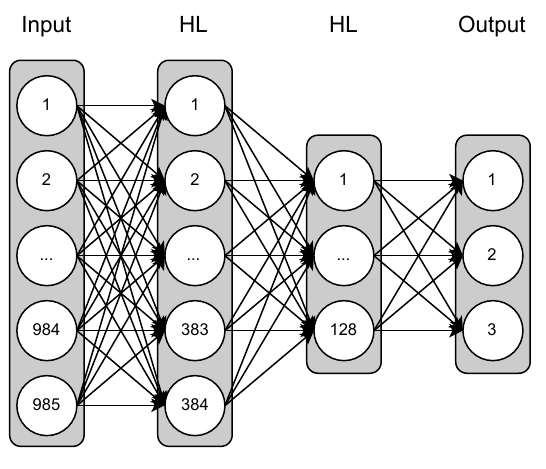}
    \caption{Neural network architecture of the calibration model as found by the Hyperparameter Optimization.}
	\label{fig:workflow_nn}
\end{figure}

As the reference model for the regression, the highly sophisticated empirical CHAOS-7 geomagnetic field model is used, consisting of different parts for the core, crustal, and magnetospheric magnetic field \cite{finlay2020chaos}.
Using all portions of the model, the model is evaluated for the present positions and timestamps of the acquired measurements from the satellite mission at hand, acting as the ground truth for the ML model.
The ML model is a feed-forward neural network (NN) with several adjustments for the training and preprocessing of the data, further details can be found in \cite{styp2022machine}.
As the reference model acting as the ground truth does not include small-scale and highly fluctuating phenomena in high latitudes, the samples are weighted according to the quality of the ground truth, with a decreasing weight towards higher latitudes.
In addition, Exponential Linear Units (ELU) and a step-wise decaying learning rate function are used \cite{clevert2015fast}.
The architecture depicted in Figure \ref{fig:workflow_nn} was found by Hyperparameter Optimization utilizing the Bayesian Optimization algorithm and consists of 384 neurons in the first hidden layer and 128 neurons in the second hidden layer. Finally, three neurons representing the 3-dimensional output \cite{snoek2012practical}.

The ML pipeline then trains a global model on all available, preprocessed, and preselected data of the mission that describes general dependencies between the input features and the measured artificial disturbances.
Subsequently, this global model is fine-tuned on a monthly basis since instruments degrade over time and calibration parameters are expected to change slightly as a function of time.
As a trade-off to having sufficient training data, a monthly scale was chosen to fine-tune the global model.

Finally, the fine-tuned models for every month are used to predict the entire initially available dataset measured by the satellite, adding flags to filter out data like outliers or unused data during the calibration, thus generating a dataset of calibrated measurements.
The monthly models generate files for every day of the corresponding month in the Common Data Format (CDF), developed by NASA\footnote{\href{https://cdf.gsfc.nasa.gov/}{https://cdf.gsfc.nasa.gov/}}.
After careful and extensive evaluation, the dataset is then published to the geoscientific community, as was also done for other satellites \cite{styp2021calibration}.

To sum up, we can extract additional information from the same set of sensors mounted on non-dedicated satellites, rendering these missions even more valuable and thus enlarging existing geomagnetic datasets without the need for additional satellite missions.
Evaluation in previous studies has shown that this calibration approach is able to greatly reduce the residuals compared to the reference model, to values of about 6.5nT for low- and mid-latitudes, allowing scientific application of the calibrated data.

    \section{Workflow Implementation}\label{sec:WORKFLOW_IMPL}
    \begin{figure}[]
\centering
    \includegraphics[width=1\columnwidth]{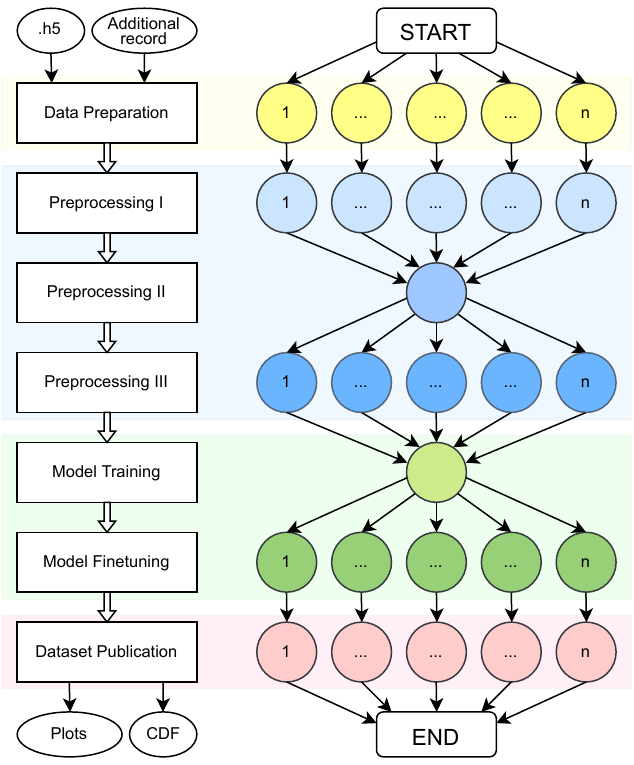}
    \caption{Workflow overview with four phases and seven abstract tasks. Section~\ref{sec:WORKFLOW_IMPL} provides a description of each phase and details on the tasks.}
	\label{fig:execution_model}
\end{figure}

This section gives insights into our workflow implementation.
Further, we explain the changes needed to transform the scripts of scientists to enable the parallelization of a scientific workflow system offers.

As a first general step to provide reproducibility and to enable our workflow to run on different kinds of systems, we created a Docker\footnote{\href{https://docker.com}{https://docker.com}} container.
The container contains all needed applications and software dependencies, e.g., python modules, Space Physics Data Facility tools by NASA, or PROJ for cartographic projections.

Figure ~\ref{fig:execution_model} provides an overview of our workflow implementation.
Our workflow consists of four phases, data preparation, preprocessing, model training, and data publication, with a total number of seven abstract tasks.

\subsection{Data Preparation}

In the first phase, the data for the analysis process needs to be prepared.
The satellite mission delivers various types of data on a monthly basis, e.g., raw magnetometer measurements, housekeeping data, and telemetry data.
Therefore, the different data sources need to be merged on a monthly basis.
While the original script looped over all available months to perform the data preparation monthly, our new abstract workflow task does this only for a single month.
This change enables the parallel execution of the script on a monthly basis, leading to a possibly higher level of parallelization since the script can be executed independently for every month.

\subsection{Preprocessing}

Filtering outliers, normalizing features, or eliminating features without sufficient values was originally conducted in a single script with the aim of preparing the model training input data.
This was necessary since some steps, like normalizing or standardizing, need aggregated counts over all months, e.g., mean, standard deviation, or counts of values.
However, to profit from a higher level of parallelization by executing the analysis on a monthly basis, we split this phase into three tasks.

\paragraph{Preprocessing I}

In the first preprocessing task, we calculate key values on a monthly basis.
These key values include means over all columns, counts, and sums.
Due to the changed structure, this task can now be executed in parallel, where the number of task instances is equal to the analyzed months.

\paragraph{Preprocessing II}

The second preprocessing phase merges the key values from all months, e.g., calculating the weighted means over all months and for each feature based on the received means and counts.
As an output, the features to drop are determined as well as the weighted mean for each feature. 
Since this task performs a merge operation, the workflow waits at this point until the first preprocessing phase is completed for all months.

\paragraph{Preprocessing III}

The third preprocessing step applies the previously determined outputs to each month's data.
This includes, for example, omitting features based on the identified obsolete features or filling gaps in the satellite measurement data based on weighted averages.
Therefore, for each month, cleaned measurement data is generated.
Again, this task is conducted in parallel, where the number of instances $n$ is determined by the months to analyze.

\subsection{Model Training}

The model training step combines the model training of the neural network and the fine-tuning of this model on a monthly basis.

\paragraph{Model Training}

In the model training task, a feed-forward neural network is trained based on all preprocessing data from the previous step and is hardly parallelizable.
One issue we encountered when testing our workflow was a huge memory peak before the training started.
A peak memory usage of 500 GB is no problem for a HPC cluster interconnected via InfiniBand but should be avoided if the computation happens on a commodity cluster since the memory of a single machine is usually much smaller.
Therefore, we had to refactor this part of our code to increase memory efficiency.
Refactoring included optimizing the used data types, preventing methods that create a data copy in memory before performing the operation, and releasing unused variables.
Besides this, no other changes were required.

\paragraph{Model Finetuning}

The model finetuning takes the output from the globally trained model together with the preprocessed measurement data on a monthly basis.
Then, an ML model is trained based on these input features for each month in parallel.

\subsection{Dataset Publication}

The monthly models from the finetuning tasks are then used to generate monthly CDF files. 
Again, the initial script did this for each month, looping over the list of available months.
In our workflow implementation, the script can be called on a single month, leading to a possible parallel execution of all task instances.

    \section{Evaluation}\label{sec:EVAL}

The first part explains our experimental setup, the used input data from the satellite systems, and the compute infrastructures.
The second part contains our two conducted experiments.

For the first experiment, we execute our workflow implementation on a commodity cluster setup.
Here, we compare the scalability of the tasks of the workflow through restricting the number of cluster nodes. 

In the second experiment, we run our developed workflow on the full commodity cluster and compare the overall makespan (time between the start of the first task and the end of the last task), task runtimes, and the resource usage on a task and workflow level.

\subsection{Experimental Setup}
\paragraph{Input Data}

The input data consists of as much information as possible about the satellite as a system. This includes the publicly available datasets of the magnetometer measurement, housekeeping data, as well as telemetry data recorded on the satellite. Together with the CHAOS-7 geomagnetic field model as the reference model, a complete dataset is generated with aligned timestamps. After the data preparation, there is a tabular dataset consisting of unique timestamps with positional information and the corresponding measurements and information from the satellite as well as the reference model. 

The magnetometer measurement, housekeeping, and telemetry data are structured monthly.
Currently, there are 46 months available from the satellite mission.

After the calibration, the resulting calibration models are applied to the whole initial dataset without any filtering or masking. 
Problematic or held out data is flagged accordingly. 
Then, a final dataset is published within daily CDF-files, named accordingly.

\begin{table}[]
\centering
\caption{Infrastructure setup overview.}
\begin{tabular}{|l|r|r|r|r|}
\hline
\textbf{}            & \textbf{Nodes} & \textbf{Threads} & \textbf{Memory} & \textbf{Storage} \\ \hline
\textbf{HPC cluster}         & 264              &   14984                  &  71976 GB                 &  $\geq$193 TB                   \\ \hline
\textbf{Commodity cluster}     & 8              & 256                 & 1024 GB             & 7 TiB               \\ \hline
\end{tabular}
\label{tab:infrastructure_cluster}
\end{table}
\paragraph{HPC cluster}

Our HPC cluster consists of 264 nodes in total consisting of 20 different kind of nodes with an age between ten and two years, as can be seen in Table \ref{tab:infrastructure_cluster}. 
This configuration leads to a total number of 14,984 threads and 71,976 GB of memory in the HPC cluster.
The nodes are interconnected via InfiniBand\footnote{\href{https://infinibandta.org/}{https://infinibandta.org/}} with different speed rates (QDR - 32 Gbps, FDR - 54 Gbps, EDR - 100 Gbps).
Slurm~\cite{yoo2003slurm} is the responsible resource manager for the cluster and allows to restrict the execution of the workload on several or single node groups.
To enable reproducible results, we restrict the execution of our original workflow on a single node group.
The node group consists of modern Intel Xeon Gold 6230 CPUs and uses Infiniband EDR with a throughput of up to 100 Gbps.
Each node consists of 2 sockets with 20 cores per socket and a total amount of 768GB DDR4 memory.

\paragraph{Commodity Cluster}
We set up a commodity cluster consisting of 8 homogeneous nodes in total, as can be seen in Table \ref{tab:infrastructure_cluster}.
Each of the nodes is equipped with an AMD EPYC 7282 16-Core Processor (32 Threads), 128GB DDR4 memory, and two 960GB SATA III SSDs.
Further, every node uses a 10 Gbps network interface for internal communication.
As the cluster manager, we installed Kubernetes (v1.23.6) together with Ceph\footnote{\href{https://ceph.io}{https://ceph.io}} (v16.2.7) for storing the files in a distributed manner.
Each node uses one disk for local storage and provides the other one for Ceph storage, leading to a total cluster storage capacity of 7 TiB.
Additionally, our cluster runs Prometheus and Grafana for monitoring the workload.

\subsection{Scalability}

\begin{figure*}
    \centering
    \includegraphics[width=\textwidth]{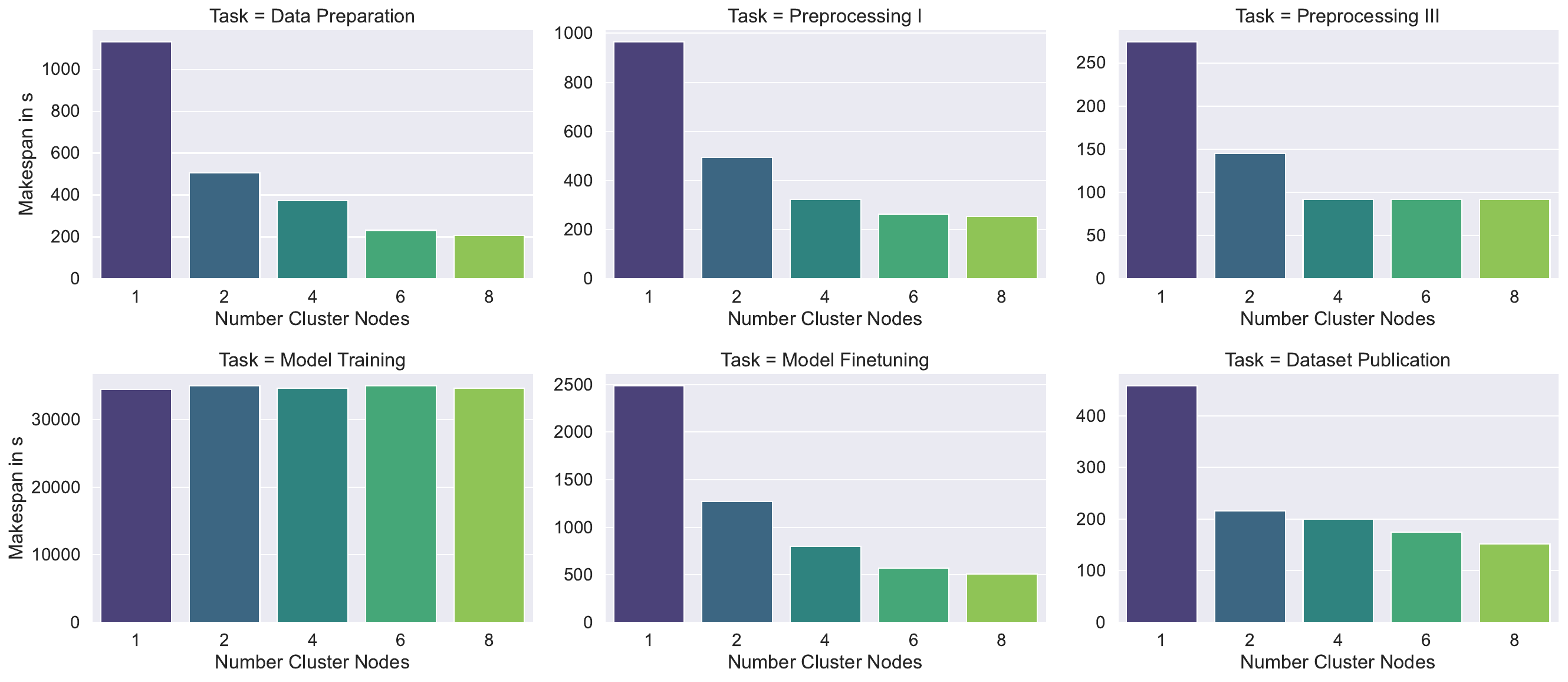}
    \caption{Runtimes of the workflow tasks on the commodity cluster with a different number of cluster nodes.} 
\label{fig:workflow_scale}
\end{figure*}

In the scalability experiment, we run our workflow on five different cluster configurations.
We start with all eight cluster nodes and remove 2 nodes in each subsequent step until there is only one node left.

Figure~\ref{fig:workflow_scale} shows the makespan (time difference between the earliest starting point and the latest endpoint of the task instances) for the six abstract workflow tasks for the different cluster configurations.
We exclude the preprocessing II task from the figure since the single task is running for an extremely short time ($<$10s) and is not dependant on the number of nodes.
The data preparation task scales well with the number of nodes, especially for up to 6 nodes.
Providing 6 nodes nearly satisfies the requirements and adding more nodes improves the runtime only slightly.
The model preprocessing I, the dataset publication, and the model finetuning task yield a similar scaling characteristic. 

An exception from the previously observed pattern is the preprocessing III task where 4 nodes are sufficient to fulfill the resource needs of this task.
Contrary, the model training task does not profit at all from an increase in the number of nodes.

Since the original implementation is conducted on an HPC cluster, where the nodes are connected via Infiniband, there is no scalability aspect to consider.

\subsection{Resource Usage}

In this section, we first compare the results of running our workflow implementation on the commodity cluster with running the original scripts on the HPC cluster.
Second, we examine the resource usages of the task instances for the abstract workflow tasks.

\begin{table*}[thpb]
\centering
\caption{The table compares the originally called scripts running on the HPC cluster with our workflow implementation running on eight commodity cluster nodes regarding runtime and allocated CPU and memory hours.}
\begin{tabular}{|l|rr|rr|rr|}
\hline
                    & \multicolumn{2}{c|}{CPU Hours Allocated}                                                  & \multicolumn{2}{c|}{Memory Hours Allocated}                                               & \multicolumn{2}{c|}{Runtime in s}                                                         \\ \hline
                    & \multicolumn{1}{c|}{Original HPC} & \multicolumn{1}{c|}{\cellcolor[HTML]{EFEFEF}Workflow} & \multicolumn{1}{c|}{Original HPC} & \multicolumn{1}{c|}{\cellcolor[HTML]{EFEFEF}Workflow} & \multicolumn{1}{c|}{Original HPC} & \multicolumn{1}{c|}{\cellcolor[HTML]{EFEFEF}Workflow} \\ \hline
Data Preparation    & \multicolumn{1}{r|}{108.24}        & \cellcolor[HTML]{EFEFEF}2.22                          & \multicolumn{1}{r|}{216.48}        & \cellcolor[HTML]{EFEFEF}26.59                         & \multicolumn{1}{r|}{12177}        & \cellcolor[HTML]{EFEFEF}218                           \\ \hline
Preprocessing I     & \multicolumn{1}{r|}{6.57}          & \cellcolor[HTML]{EFEFEF}0.41                          & \multicolumn{1}{r|}{102.64}        & \cellcolor[HTML]{EFEFEF}4.94                          & \multicolumn{1}{r|}{739}          & \cellcolor[HTML]{EFEFEF}238                           \\ \hline
Preprocessing II    & \multicolumn{1}{r|}{-}            & \cellcolor[HTML]{EFEFEF}0.00                          & \multicolumn{1}{r|}{-}            & \cellcolor[HTML]{EFEFEF}0.00                          & \multicolumn{1}{r|}{-}            & \cellcolor[HTML]{EFEFEF}6                             \\ \hline
Preprocessing III   & \multicolumn{1}{r|}{8.34}          & \cellcolor[HTML]{EFEFEF}0.44                          & \multicolumn{1}{r|}{68.79}         & \cellcolor[HTML]{EFEFEF}6.98                          & \multicolumn{1}{r|}{938}          & \cellcolor[HTML]{EFEFEF}91                            \\ \hline
Model Training      & \multicolumn{1}{r|}{341.14}        & \cellcolor[HTML]{EFEFEF}293.18                        & \multicolumn{1}{r|}{2046.83}       & \cellcolor[HTML]{EFEFEF}1172.70                       & \multicolumn{1}{r|}{19189}        & \cellcolor[HTML]{EFEFEF}34048                         \\ \hline
Model Finetuning    & \multicolumn{1}{r|}{123.82}        & \cellcolor[HTML]{EFEFEF}16.88                         & \multicolumn{1}{r|}{371.47}        & \cellcolor[HTML]{EFEFEF}22.50                         & \multicolumn{1}{r|}{6965}         & \cellcolor[HTML]{EFEFEF}401                           \\ \hline
Dataset Publication & \multicolumn{1}{r|}{42.65}         & \cellcolor[HTML]{EFEFEF}1.32                          & \multicolumn{1}{r|}{255.89}        & \cellcolor[HTML]{EFEFEF}6.58                          & \multicolumn{1}{r|}{2399}         & \cellcolor[HTML]{EFEFEF}325                           \\ \hline
Total (Makespan)              & \multicolumn{1}{r|}{630.76}        & \cellcolor[HTML]{EFEFEF}314.43                        & \multicolumn{1}{r|}{3062.09}       & \cellcolor[HTML]{EFEFEF}1240.30                       & \multicolumn{1}{r|}{42407}        & \cellcolor[HTML]{EFEFEF}34972                         \\ \hline
\end{tabular}
\label{tab:results}
\end{table*}
Table ~\ref{tab:results} compares the allocated CPU hours, the allocated memory hours, and the runtime of the original HPC scripts with our workflow.
The allocated hours $h_{alloc}$ are defined as:
\begin{equation}
h_{alloc} = \sum_{n=1}^{m}  t_n \cdot alloc 
\end{equation}
where $m$ is the number of task instances, $t_n$ is the time the specific task instance $n$ run in hours, and $alloc$ the resources (CPU cores or Memory in GB) allocated to the task.
The preprocessing II task is not available for the original HPC values since we newly introduced this task to enable 	parallelism in our workflow. 

In total, our workflow implementation reduced the total number of CPU hours by 316.37 hours, a decrease of 50.2\%.
The biggest decrease can be observed for the data preparation task, where the original workflow allocated 108.20 CPU hours while our workflow implementation only used 2.22 CPU hours.
The smallest observable difference yields the model training task where the workflow achieved a reduction of 14.1\%.

Regarding allocated memory hours, we achieved a reduction of 59.5\% compared to the original implementation.
The data publication task yields the biggest percentage change of 97.4\%.
Again, the smallest percentage reduction is observable for the model training task, however, this is still a reduction of 42.7\%.

Lastly, the workflow implementation decreased the overall makespan of the workflow by 17.5\%.
Thereby, the model training task accounts for 97.4\% of the total time of the workflow.
Contrary, the original implementation spends only  45.3\% of the time on the model training task.
Therewith, the model training task takes significantly longer in the workflow implementation. 
During all other tasks, the workflow is able to reduce the runtimes.

\begin{figure}[!h]
\centering
    \includegraphics[width=1\columnwidth]{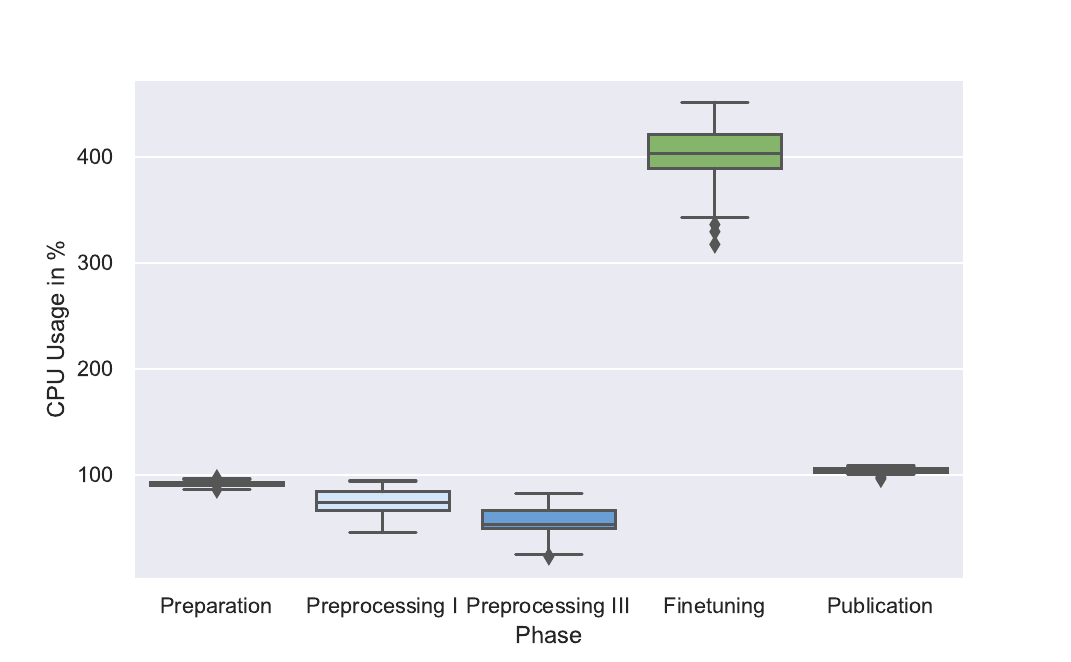}
    \caption{CPU task usages in percentage over all task instances when using eight cluster nodes.}
	\label{fig:workflow_cpu}
\end{figure}

\begin{figure}[!h]
\centering
    \includegraphics[width=1\columnwidth]{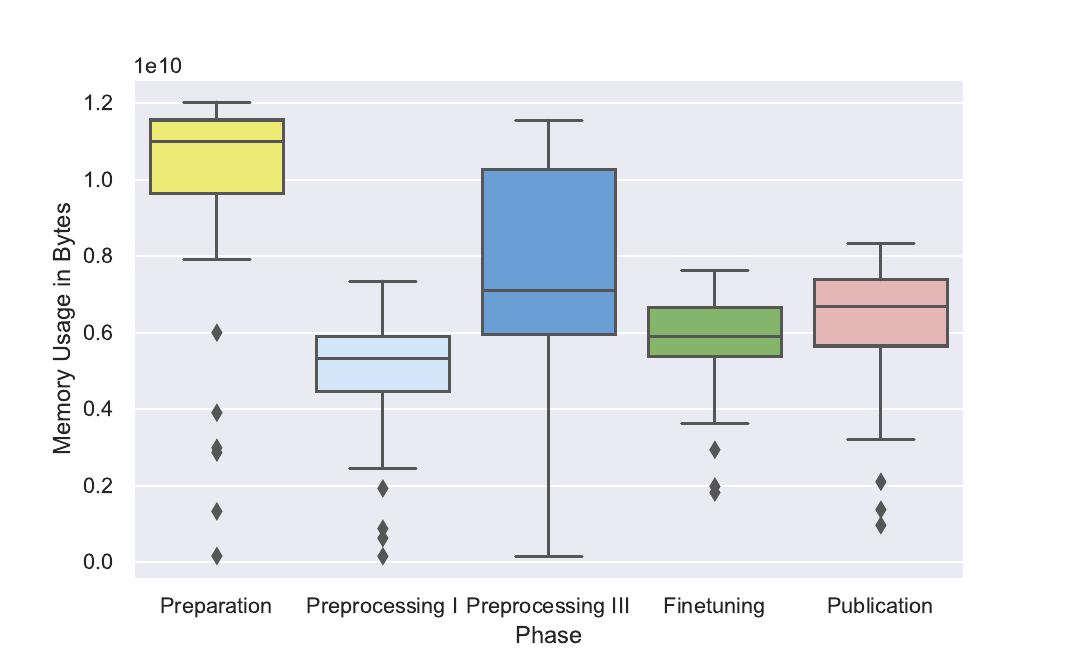}
    \caption{Memory task usages in bytes over all task instances when using eight cluster nodes.}
	\label{fig:workflow_mem}
\end{figure}

Figure~\ref{fig:workflow_cpu} shows the CPU usages of the tasks in percent, where 100\% is equal to one used core.
As mentioned before, each task in the workflow is $n$ times executed, where $n$ is the number of observed satellite months.
Exceptions are the preprocessing II and the model training task that merge the incoming results and are only executed once each.
Therefore, they are excluded from Figure~\ref{fig:workflow_cpu} and Figure~\ref{fig:workflow_mem}.
One can see that the cpu usages of the tasks are rather stable and yield a low standard deviation. 
One exception from this observation is the finetuning task where a higher standard deviation can be seen.
In general, four of the five tasks only use one CPU core and do not profit from more CPUs assigned.

Figure~\ref{fig:workflow_mem} visualizes the memory consumption of the tasks in bytes and shows a different pattern.
Here, a higher relative standard deviation can be seen.
Further, one can observe significant outliers, e.g., the median memory consumption of the data preparation task is 10.5 GB, however, one instance only consumed 173.4 MB.
The preprocessing II task yields the highest range between upper and lower whisker.

Overall, the evaluation shows that implementing this scientific pipeline into a scientific workflow yields significant impact.
After the full implementation, the workflow was able to decrease the allocated CPU hours by 50.2\%, the allocated memory hours by 59.5\%, and the makespan by 17.5\%.

    \section{Discussion}\label{sec:DISCUSSION}
    Summarizing, our evaluation results indicate that switching from a traditional approach where scripts are sequentially called to a workflow implementation yields several advantages.

First, scientific workflow management systems provide functions to implicitly parallelize the execution of a task.
For existing code like ours, this can usually be achieved by changing the script input, e.g., processing on a monthly basis and merging the results later.

Second, through the parallelization, the workflow can also run well on a commodity cluster where the nodes are possibly weaker since they are not interconnected. 
Table~\ref{tab:results} shows that we were able to achieve a lower makespan.
However, tasks that are hardly parallelizable might suffer from an execution on such a commodity cluster due to memory or CPU restrictions, e.g., the model training task.
In the original implementation, the model training task accounted for around 45.3\% of the makespan time.
However, in our workflow implementation, it accounted for 97.4\% of the makespan time.
In this specific example, this is likely due to the implementation where the training of the neural network is restricted to one machine (or multiple interconnected machines via a very fast link, e.g., Infiniband).
More precisely, the machine learning task executed on the commodity cluster was restricted to 32 threads and 128GB of memory (the configuration of a single node), whereas the HPC execution with the original script had access to more than 64 threads and 384 GB of memory.
Distributed model learning approaches could circumvent the restriction on a single node in the future.

Third, this parallelization enables the use of more fine-granular resource settings.
This can be helpful to avoid resource wastage since Figure~\ref{fig:workflow_cpu} pointed out that many tasks are not able to use many CPU cores, e.g., due to the underlying framework or the implementation.
Further, our evaluation showed that a task's memory consumption can vary highly.
This is a frequent pattern since related work indicates a dependency between input data size and memory usage or runtime~\cite{witt2019feedback, baderLotaruLocallyEstimating2022, will2022get}.
In our workflow, this is due to months with more or significantly less satellite measurement data which can originate in satellite outages or data gaps.

    \section{Conclusion}\label{sec:CONCLUSION}
    This paper presented Macaw, our machine learning magnetometer calibration workflow.
We described our workflow in detail and explained the single steps with the implementation changes needed to port our old sequentially executed scripts to our Macaw workflow.

In our experimental evaluation, we ran our workflow implementation on a commodity cluster and compared it with the execution of the original scripts on a HPC cluster.
In total, our workflow decreased the overall makespan by 17.5\%.
Even more significant is the reduction in allocated resources, where Macaw reduced the allocated CPU hours by 50.2\% and the memory hours by 59.5\%.

Future work may include enabling the parallel execution of hardly parallelizable machine learning tasks like model training on multiple cluster nodes by using federated learning.

\section*{Acknowledgments}
    \thanks{
    Funded by the Deutsche Forschungsgemeinschaft (DFG, German Research Foundation) as FONDA (Project 414984028, SFB 1404) and HEIBRIDS - Helmholtz Einstein International Berlin Research School in Data Science under contract no. HIDSS-0001.
    }
    \bibliographystyle{IEEEtran}
    \balance
    \bibliography{./references}

\end{document}